\documentclass[%
 reprint,
 amsmath,amssymb,
 aps,
]{revtex4-1}

\usepackage{graphicx}
\usepackage{dcolumn}
\usepackage{bm}
\usepackage{topcapt}

\begin{document}

\preprint{APS/123-QED}

\title{Casimir forces and high-Tc superconductors}

\author{Carlos Villarreal and Santiago F. Caballero-Benitez}
 
\email{scaballero@fisica.unam.mx}
\affiliation{Instituto de F\'isica, Universidad Nacional Aut\'onoma de M\'exico, Ciudad Universitaria, Mexico City, 04510, Mexico\\
}%


\begin{abstract}
We investigate the Casimir forces between high-$T_c$ superconductors as function of the distance and temperature, focusing on optimally-doped  YBa$_2$Cu$_3$O$_{6.95}$. We consider formerly studied configurations in normal metals lying in the  short-distance (50 nm - 600 nm), and long-distance (600 nm - 10 microns) regimes.  The dielectric properties of the material are described in terms of weakly-interacting, short-correlated conducting pairs transported along quasi-2D layers.  In the short-range regime, a continuous behavior of the Casimir forces arises, with no significant discontinuity at the transition temperature. This behavior also follows in the equivalent the normal conductor configuration. In the long-range regime, the forces show an abrupt increment at the critical temperature. Simultaneously, the Casimir entropy and the specific heat develop a strong discontinuous behavior, characteristic of the SC-normal phase transition. In every situation, the entropy vanishes at extremely low temperatures.
\end{abstract}

\maketitle


\emph{Introduction.}
Casimir forces are induced by the distortion of the spectrum of quantum and thermal fluctuations of the electromagnetic field in presence of material
bodies. The original theory proposed by Casimir in 1948 \cite{Casimir48} predicts that two parallel perfectly-conducting surfaces separated by a distance $d$ will be subject to an attractive force per unit area $F_C=-\hbar c/240 d^4$.  Based on the fluctuation-dissipation mechanism, in 1956 Lifshitz put forth a more realistic approach that takes into account the dielectric properties of materials \cite{Lifshitz56}.  In the last twenty years numerous experiments \citep{Lamoreaux97,Harris00,Bressi02,Decca05,Jourdan09,Chang13,Banishev13, Castillo13} have been performed  on measuring the Casimir forces in a wide diversity of experimental setups  \citep{Bordag09,Klimchitskaya09,Rodriguez11,Klimchitskaya11}. In these experiments, the observed data are compared with theoretical predictions arising from Lifshitz theory, which requires the knowledge of the optical response of materials in a broad frequency range. 
Unexpectedly, the low-frequency behavior of the complex permittivity has prompted an intense theoretical debate related with the consistency of Lifshitz theory with Nernst's principle (the third law of thermodynamics) \cite{Milton04,Bezerra04,Geyer05,Brevik05,Bezerra06,Mostepanenko10}. In the low-frequency regime, the optical response of materials with dispersive and absorbing properties is typically described by means of the complex Drude dielctric function, where the dissipative transport of conduction electrons is accounted for by a relaxation rate $\gamma_0$. For metals where the charge carriers constitute a tenuous electronic plasma,  damping is negligible ($\gamma_0 \to 0$) and the Drude model is well approximated by the plasma  dielectric function.

In principle, the inclusion of dissipative contributions in the permittivity should be a necessary condition to achieve formal congruence with the fluctuation-dissipation relation. Furthermore, the consideration of this mechanism is essential to describe quantum properties of absorbing media, such as the preservation of commutation relations inherent to quantum atomic radiators constituting a given material \citep{Milloni94}. A number of measurements of Casimir forces in metals at room temperature and involving relatively small body separations (50 - 600 nm) have been performed with better predictions based on the plasma rather than the Drude model, but the relative difference between either model predictions is tiny. 
 Up to date, a few experiments carried out at large separations (600 nm - 8 $\mu$m) show better agreement with theoretical predictions including electronic relaxation \cite{Sushkov11,Garcia12}. 
 The origin of this discrepancy is the influence of thermal electromagnetic fluctuations, which at micrometers predominate over zero-point fluctuations, so that $k_B T \gg \hbar \omega_c$, $\omega_c=c/2d$, and $c$ is the speed of light.

The thermodynamic aspects of the Casimir effect have motivated investigations on the detailed temperature dependence of the force \cite{Bostrom00,Bostrom04, Brevik05,Brevik14}. Measurements of the Casimir force between Au surfaces at 300, 77, 4.2, and 2.1 K  \cite{Decca10} show that while low-temperature results are noisier than room temperature ones, the average of the measurements coincides at all temperatures. This suggests that thermal fluctuations play a negligible role in these experiments. Therefore, this precludes a direct exclusion of either model Drude or plasma. Similar conclusions are derived from measurements of the normalized gradient of the Casimir force at 77 K with liquid nitrogen \cite{Castillo13}.
In this context, superconducting (SC) materials have been suggested as an ideal scenario to study this problem. Theoretical studies based on standard BCS superconductors such as Ni, Al, or NbTiN, with a transition temperature $T_c \approx 1 - 10$ K have been made \cite{Bimonte05,Bimonte19}. Nevertheless, the predicted effects associated to the SC transition are inconclusive with current experimental techniques. This agrees with recent experiments probing the variation of the Casimir force between two closely spaced thin Al films at $T_c$, observing a null result\cite{Norte}. 

In this work we study the Casimir forces between high-Tc superconductors (HTSCs) where vacuum and thermal fluctuations can be comparable in some regimes. We focus on the optimally-doped ceramics YBa$_2$Cu$_3$O$_{6.95}$ (YBCO), with $T_c= 93$ K \citep{Wu87} and perform a thorough exploration of distance and temperature regimes involved in experimentally feasible setups. We considered two alternative configurations associated to  short- and long-range regimes. They both include a planar surface separated by a distance $d$ from a spherical surface with curvature radius $R$, covered both by an YBCO film. In the short-range regime, the setup involves a sphere with  $R = 95 \mu$m, while the long-range it involves a spherical lens with $R = 15$ cm. Surprisingly, we find that at long distances it is possible to observe the effects related to superconductivity clearly. In order to compare with results expected from ordinary non-SC materials, we also contemplated an artificial material with the same optical properties as YBCO, except  that $\gamma_0$ remains finite even at $T<T_c$. In the following, this material is denoted as ordinary conductor.
 
\emph{Model: Lifshitz formulation.}
The Casimir force between a slab and a spherical surface of curvature radius $R$ may be evaluated by means of the Derjaguin approximation \citep{Milloni94}, valid in the limit $R\gg d$: $F(d) \simeq 2\pi R {\cal F}_\parallel (d)$, where ${\cal F}_\parallel(d)$ is the free energy per unit area between two $parallel$ surfaces. In the reflection coefficients formulation of Lifshitz theory \cite{Lambrecht00,Mochan06}, the expression for free energy density  at finite temperature is obtained by considering the Matsubara formalism with discrete imaginary frequencies $ \omega_n= 2\pi i n k_B T/\hbar$, and integer $n$:  
\begin{equation}
 {\cal F}_\parallel (d)=k_B T {\sum_{n=0}^{\infty}} ^\prime \int_0^{\infty} \frac{d^2 k_\perp}{(2\pi)^2} \sum_{\alpha=s,p} \ln\left[1-r_\alpha^2 e^{-2 \kappa_0 d} \right] , 
\end{equation}
where the prime in the $n$-summation means that the $n = 0$ term must be halved. Here, $k _\perp$ is the wave vector component parallel to the plates,  $\kappa_j=\sqrt{k_\perp^2 + \epsilon_j \omega_n^2/c^2}$, $\alpha$ the field polarization, and  $r_\alpha\left[ \epsilon_j(i\omega_n)  \right]$ the electromagnetic reflection coefficients. The subindex in the dielectric function $\epsilon_j$ refers to the medium between the plates ($j=0$) and the plates themselves ($j=1$).  The reflection coefficients for $s$ and $p$ polarizations are given by
$r^{s}=\left(\kappa_1-\kappa_0\right)/\left(\kappa_1+\kappa_0\right)$, $r^{p}=\left(\epsilon_0 \kappa_1-\epsilon_1\kappa_0\right)/\left(\epsilon_0\kappa_1+\epsilon_1\kappa_0\right)$.
The results provided by Lifshitz theory must be yet corrected by the existence of  roughness in the surfaces involved in an experiment.
This can be simply modeled by assuming a stochastic variation in the surface separation with a root mean squared amplitude
$A_{rms}$. A Taylor expansion then yields  $ F_C(d) \simeq F(d)\left(1+6 A_{rms}^2/d^2 + ...\right)$ \cite{Harris00}.

\emph{Effective model for HTSCs.} The optical properties of HTSCs have been experimentally investigated
for different compounds at several temperatures and frequencies using reflectivity and impedance-type measurements \cite{Timusk88a,Timusk88b,Timusk05}. In the case of YBCO, the measured optical response has been represented in terms of a  dielectric function $\varepsilon(\omega)$ with parameters estimated in the normal and SC regimes  at $T=100$K, and $T=2$K, respectively. This description is in agreement with the London's two-fluid model of superconductivity \cite{Annet04}, which assumes that the number density of charge carriers $n$ may be divided into normal $n_n(T)$, and superfluid $n_s(T)$ fractions, such that $n=n_n(T)+n_s(T)$. This allows the introduction of corresponding plasma frequencies $\omega_{pn}(T) \sim n_n(T)^{1/2}$ and $\omega_{ps}(T) \sim n_s(T)^{1/2}$ (or the associated  magnetic penetration depth $\lambda_p(T)=c/\omega_{ps}(T)$). The former considerations are integrated into a dielectric function including intra-band,  Drude, mid-infrared, and optical phonon contributions. In the normal state, these are given by:
\begin{eqnarray}
\varepsilon_n(\omega)=\varepsilon_{\infty}&-&\frac{\omega_{pn}^2(T)}{\omega^2 + i \gamma_0 \omega} - \frac{S_{ir}  \omega_{ir}^2}{\omega^2 -\omega_{ir}^2 + 
i \gamma_{ir} \omega}  \nonumber \\
& &-\sum_{l=1}^{N_{ph}} \frac{S_l  \omega_{ph,l}^2}{\omega^2-\omega_{ph,l}^2+  i\gamma_{ph,l}  \omega}.
\end{eqnarray}
Here, $\varepsilon_\infty = 3.8$, $\omega_{pn}(100 \mathrm{K})=0.75$ eV, and the electronic relaxation $\gamma_0=0.037$ eV.
The parameters $\omega_i$, $\gamma_i$, and $S_i$ denote the  characteristic frequencies, relaxation rates, and oscillator strengths of the rest of contributions mentioned above, and they are presented as Supplemental Material. In the following we assume that in normal state, but close to $T_c$, the plasma frequency has the fixed value $\omega_{pn}(100 \ K)\equiv \omega_{pn}$.
In the SC state, dissipative scattering does not occur ($\gamma_0 \to 0$) and the Drude contribution  collapses to a delta function at the origin:
\begin{eqnarray}
&\varepsilon_{s}&(\omega)=\varepsilon_{\infty}+\frac{ i \pi \omega_{ps}^2(T)}{2 \omega}  \delta(\omega)-\frac{\omega_{ps}^2(T)}{\omega^2 } \\ \nonumber
&-& \frac{S_{ir}  \omega_{ir}^2}{\omega^2 -\omega_{ir}^2 + i\gamma_{ir} \omega}  
-\sum_{l=1}^{N_{ph}} \frac{S_l \omega_{ph,l}^2}{\omega^2-\omega_{ph,l}^2+  i\gamma_{ph,l} \omega}.
\end{eqnarray}
As before, the parameter $\varepsilon_\infty = 3.8$, while $\omega_{ps}(2\mathrm{K})=0.75$ eV. The rest of parameters is also found as Supplemental Material. 

In order to establish an explicit expression for $\omega_{ps}(T)$, we assume that, even if the pairing mechanism in HTSCs is uncertain, their charge transport properties may be described in terms of a 2D gas of weakly-interacting fermion pairs that condense at $T_c$. The 2D character reflects the fact that HTSCs comprise layered crystallographic structures in which charge transport occurs along copper planes (CuO$_2$) \citep{Fujita03}. The rest of the premise relies on the observation that, except for very weak coupling, fermion pairs form and condense at different temperatures \citep{Chen05}. In  BCS superconductors, the weak interaction leads to simultaneous pair formation and superfluid behavior at $T_c$. In contrast, in HTSCs the strong fermionic interactions induce short-correlated pairs ($\xi_0 \sim 1$ nm) with no phase coherence already at temperatures $T^*>T > T_c$, conforming a  pseudogap (PG) phase that precedes the onset of superfluidity at $T<T_c$. It is now established \citep{Chen05, Calvanese18},  that BCS pairs and PG pairs are two limiting cases in the BCS-BEC crossover. This crossover, smoothly connects the pairing in momentum space (BCS limit), with pairing in position space (BEC limit), being a hallmark of the latter the emergence  of localized pairs approximately satisfying Bose statistics. These compound particles define a very  dilute gas in YBCO, as the ratio $\kappa=\lambda_p(0)/\xi(0) \approx 95$, so the mean interparticle  distance 

\newpage
\noindent
\onecolumngrid
\begin{figure*}[th]
\begin{center}
\includegraphics[width=0.85\textwidth]{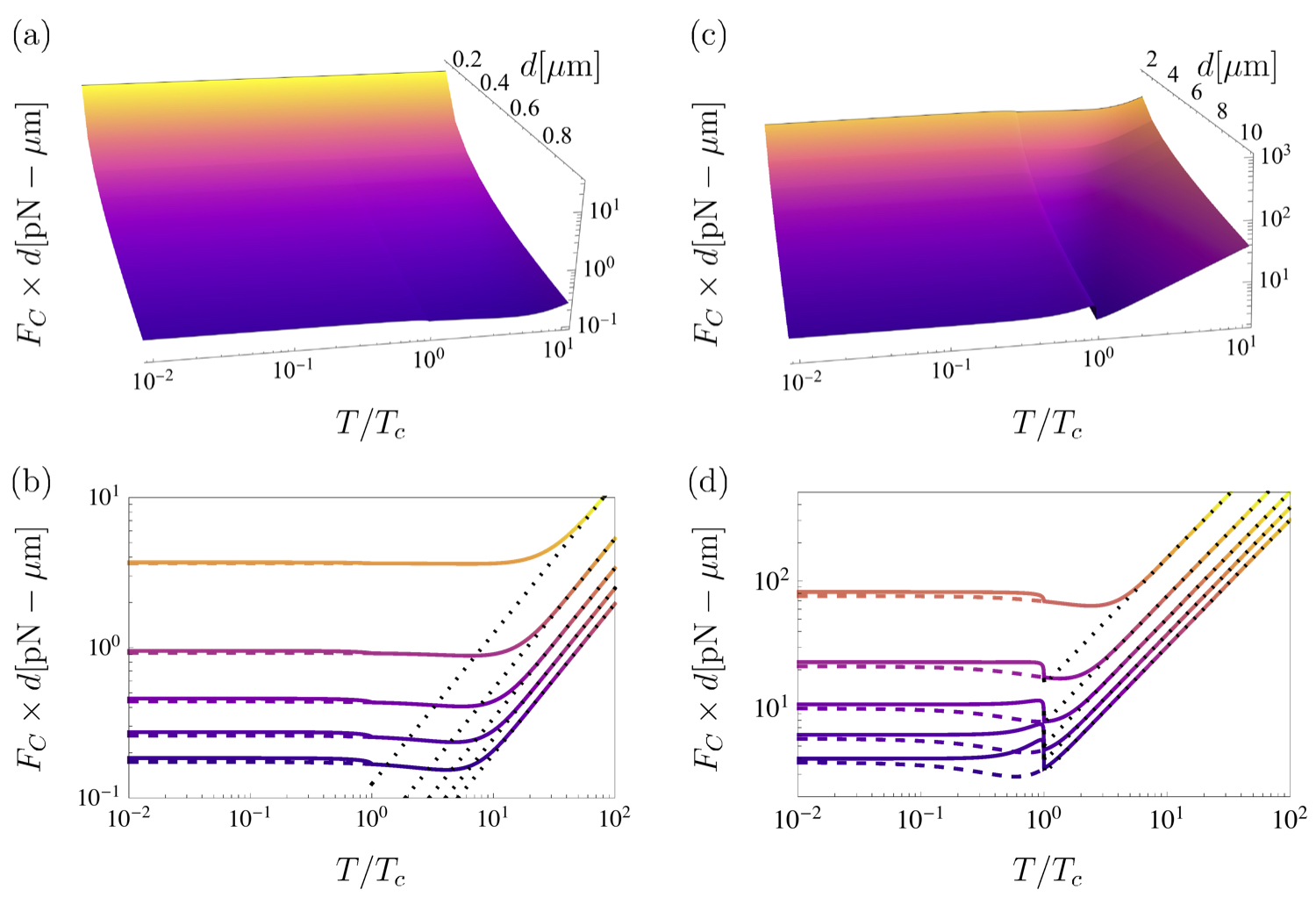}
\end{center}
\caption{
(color on-line) Casimir force (times separation) as function of the reduced temperature $T/T_c$ and distance $d$ in $\mu$m. a) Force surface for the short-range YBCO configuration with sphere radius $R=95 \mu$m. b) Projections of the force surface for fixed separation values. From the upper (yellow, bright) to the lower (purple, dark) curve: $d = 0.2, \ 0.4, \ 0.6 , \  0.8, \ 1.0 \ \mu$m.
 Dashed curves correspond to a normal conductor, while dotted curves correspond to the high-temperature limit $F_C^{\mathrm{HT}} = \zeta(3) R k_B T/ 8d$. c)
Force surface for the long-range YBCO configuration with spherical lens of curvature radius $R=15$ cm. d) Projections of the force surface for fixed separation values. From the upper (yellow, bright) to the lower (purple, dark) curve: $d = 2, \ 4, \ 6 , \  8, \ 10 \ \mu$m.
 Dashed curves correspond to a normal conductor, while dotted curves correspond, as before, to the high-temperature limit.}
\label{fig:Figure1}
\end{figure*}
\twocolumngrid
\noindent
largely exceeds the pair size.  On these grounds, the dynamics of HTSCs charge carriers may be accounted by means of a weakly-interacting Bose gas embracing elementary excitations with a Bogoliubov spectrum $\mathcal{E}(p)=\left(p^2 c_s^2+(p^2/2m)^2 \right)^{1/2}$ (with $c_s$ the sound speed); in the low-momentum limit, this yields a phonon spectrum, $\mathcal{E}(p) \to c_s p$. It is straightforward to show that a 2D gas with a linear spectrum satisfies \citep{Lomnitz13}:
\begin{equation}
 \omega_{ps}^2(T)/\omega_{ps}^2(0) =\lambda_{p}^{-2}(T)/\lambda_{p}^{-2}(0)=1-(T/T_c)^2.
\end{equation}
This relation gives an accurate representation of  experimental measurements of the penetration length $\lambda_p(T)$ in the CuO$_2$ plane for a wide range of dopings of  YBa$_2$Cu$_3$O$_{6+x}$ samples \citep{Zuev05,Chen05}, as well as of the dependence of the transition temperature on doping \citep{Broun07,Lomnitz13}. 
In conjunction with Eqs. (1)-(3), this expression provides a closed framework for the evaluation of the Casimir forces in HTSCs.  

\emph{Results and discussion.}
The calculated Casimir forces involve qualitatively distinct features in the short- and long-range configurations, as may be observed in the figure panel (1).  There, we present surfaces representing the force magnitude, $F_C(d,T)$, as well as surface cuts, $F(T)$, for fixed $d$. 
In the short-range regime depicted in Fig. (1.a)), $F_C(d,T)$ displays the expected continuous behavior as function of $d$ in almost the whole distance and temperature ranges, except for a barely discernible discontinuity at $T=T_c$ at relatively large distances $d \sim 1 \mu$m. This behavior is congruent with previous experimental and theoretical studies for setups involving normal conductors as well as BCS superconductors. Fig. (1.b) shows that for $T/T_c $ larger than an order of magnitude, the force decreases linearly with temperature, accordingly with the expected high-temperature behavior $F_C^{\mathrm{HT}}(T) =\zeta(3) R k_B T/d^2$. At smaller temperatures $F_C(T)$ develops a shallow minimum, and for temperatures $T<T_c$ the force acquires a constant value.
In the long-range regime (Figs. (1.c) and (1.d)), $F_C(d,T)$ also displays a continuous behavior as function of $d$. Surprisingly, the temperature dependence exhibits a marked discontinuity at $T=T_c$, with increasing magnitude for the largest separations, being
the maximum increment at $T<T_c$  about 70 $\%$ of its value at $T\ge T_c$. Evidently, this is correlated with the SC transition, and as we will see below, the fact that the field fluctuations and the pairs have to balance their fluctuations.
On the other hand, for temperatures $T\ll T_c$ the force shows a constant behavior, as also observed in the short-range configuration.
In the case of a normal conductor, the Casimir forces show a continuous behavior paralleling that arising in the short-range regime of HTSCs. However, this behavior is apparent now in either the short- and long-range regimes.
\begin{figure}[t]
\centering
\includegraphics[width=\linewidth]{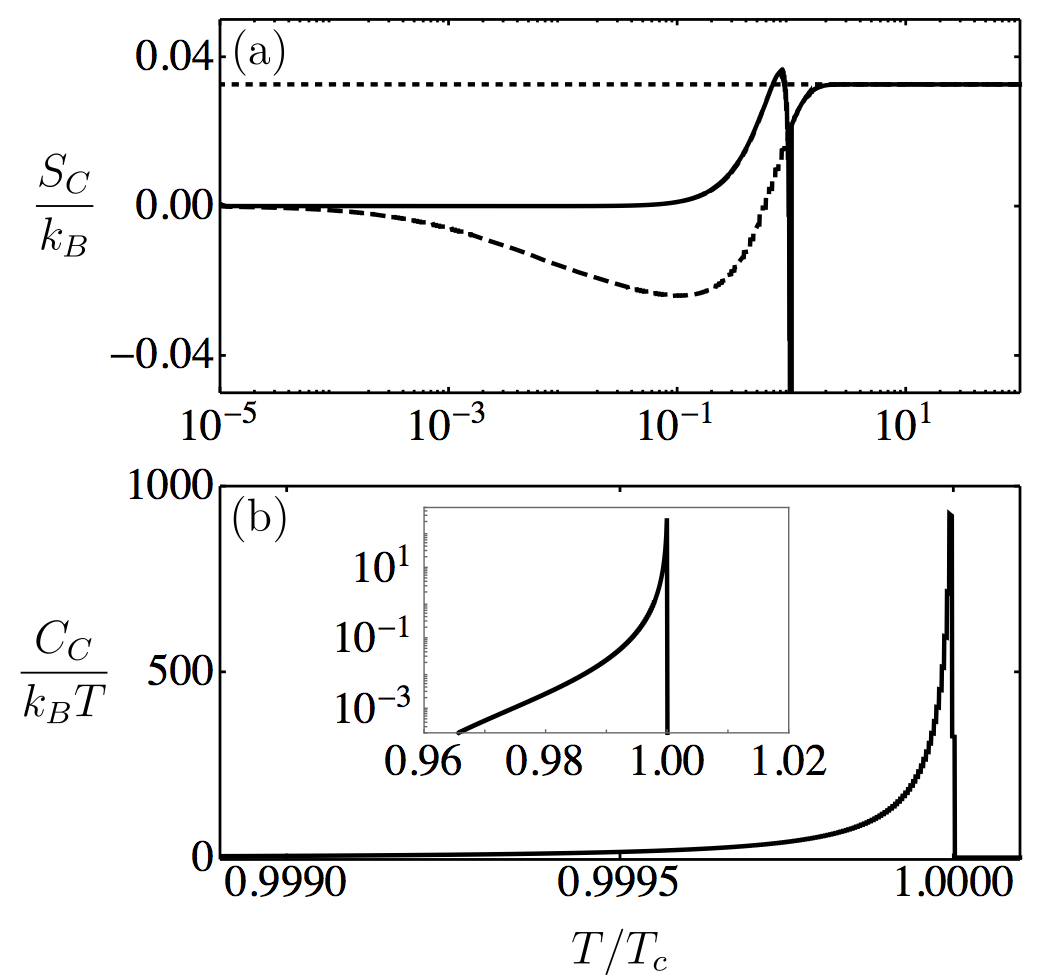}
\caption{ (a) Casimir entropy for a long-range configuration as function of the reduced temperature $T/T_c$ for a fixed separation  $d=10 \mu$m. The continuous line corresponds to a  configuration involving YBCO,  the dashed line corresponds to a normal conductor, and the dots (constant) are the high temperature limit ($S_C^{\mathrm{HT}}/k_B=\zeta(3)R/d^2$).  (b) Casimir specific heat (over temperature) for the former YBCO configuration.  The inset shows the same quantity displayed in logarithmic scale.
}
\label{fig:Figure2}
\end{figure}

Figure (2) depicts the behavior of the Casimir entropy $S_C/k_B=- \partial {\cal F}_\parallel/\partial T$ for both a HTSC and a standard conductor. The entropy develops  a strong negative divergence in the neighborhood of $T_c$, signaling the SC-Normal phase transition. On the other hand, $S_C$ acquires constant values for $T>T_c$, as well as $T\ll T_c$. Clearly, $S_C \to 0$ as $T \to 0$ in either case. The vanishing of the entropy is a necessary consequence of the constant value attained the free Casimir energy in this limit, which is in turn proportional to the Casimir force in the Derjaguin approximation.
The specific heat coefficient $\gamma=C_C/(k_B T)=\frac{\partial}{\partial T} (S_C/k_B)$ shows a concomitant sharp increase at $T_c$, relaxing steadily to a null value for $T<T_c$. Remarkably, this behavior is very similar to that observed in experimental determinations of the specific heat in YBCO \cite{Annet04,Chen05}, associated in that case to conducting pair excitations. Indeed, this would be a result of the existence of thermodynamic equilibrium between the SC material and electromagnetic field fluctuations. Thus, the field inherits the properties of the fermionic pairs.
On the other hand, in the case of standard conductors, the Casimir entropy has a very similar behavior as that predicted for Cu samples in Ref.\cite{Bostrom04}. 
It displays a constant value for $T>T_c$, steadily decreasing to a minimum negative value for $T<T_c$, and finally vanishing for extremely low temperatures even in presence of dissipative effects in the optical response of these materials. Similarly, the predicted behavior of the Casimir forces is congruent with that reported in Ref.\cite{Brevik05} by means of a Drude model for the dielectric response of Au and Al. Thus, consistency with the Nernst principle is achieved in every studied system in this work.

We conclude that measurements of the Casimir force in YBCO performed in a long-range setup should display a discernible discontinuity at the transition temperature $T=93$ K. Moreover, the interplay between the fluctuations in the pairs and the field could be investigated in this regime. These studies  would be feasible according to experimental reports presented in Refs. \citep{Decca10,Castillo13}.  Our calculations show that the associated entropy in the limit of low temperatures should vanish in all considered cases, consistently with the theoretical results obtained for normal conductors with dissipative charge transport. However, the extremely low values of the temperatures involved seem unreachable by using current experimental techniques to verify the entropy behaviour.

\begin{acknowledgements}
This work was partially funded by grants,  DGAPA-UNAM (IN109619), and  CONACYT-Mexico (A1-S-30934). We would like to thank DICu of the SMF for their hospitality and financial support in the meetings 2017 and 2018 where relevant discussions regarding this work took place.
\end{acknowledgements}
 
\bibliography{test}
\section{Supplemental Material}
\begin{table}[htbp]
\caption{Parameters of dielectric function of YBCO$_{7-\delta}$ in normal ($T=100$ K), and superconducting states ($T = 2$ K).}
\begin{tabular}{l c c c c c c c}
\hline
\hline
\ \ \ & &$T = 100$ K & & &$T = 2$ K & & \\
\ \ \ j  &$\omega_j$(cm$^{-1})$&$\gamma_j$(cm$^{-1})$&$S_j$& &$\omega_j$(cm$^{-1})$&$\gamma_j$(cm$^{-1})$&$S_j$ \ \ \ \\
\hline
\ \ \ 1&155&3.1&31& \ \ \ &155&2.4&31\\
\ \ \ 2&195&9.6&3& \ \ \ &194&2.7&2\\
\ \ \ 3&279&18&6& \ \ \ &277&21&10\\
\ \ \ 4&314&11&10& \ \ \ &311&9&12\\
\ \ \ 5&532&57&2& \ \ \ &534&50&3\\
\ \ \ 6&569&16&2& \ \ \ &567&17&2\\
\hline
\end{tabular}
   \end{table}

\end{document}